# REDUCTION DE LA NUISANCE OLFACTIVE PAR OXYDATION PHOTOCATALYTIQUE

G. Vincent, O. Zahraa

Département de Chimie Physique des Réactions (DCPR), UMR 7630 CNRS-INPL, ENSIC, 1 rue Grandville, BP 20451, 54001 Nancy Cedex (France)

**Résumé.** La photocatalyse hétérogène apparaît comme une technologie alternative efficace pour la purification, la décontamination et la désodorisation des effluents gazeux. L'équipe « Génie Photocatalytique » du DCPR s'est intéressé à l'étude de la 2-butanone ou méthyléthylcétone (MEK), Composé Organique Volatil (COV) de faible poids moléculaire (< 100 g/mol), pouvant engendrer de sérieuses nuisances olfactives. Plusieurs paramètres cinétiques comme l'influence de la concentration initiale et l'éclairement incident ont été étudiés sur la photodégradation de la MEK au sein d'un photoréacteur annulaire. Comme la plupart des COVs, la dégradation de la MEK semble suivre le modèle cinétique de Langmuir-Hinshelwood (LH). Les produits intermédiaires gazeux ont été analysés par GC/MS et l'acétaldéhyde s'est avéré comme le sous-produit majoritaire lors de la dégradation photocatalytique.

## OXYDATION PHOTOCATALYTIQUE DE LA METHYLETHYLCETONE (MEK)

### 1. Méthyléthylcétone

La MEK est un liquide incolore, volatil, dégageant une forte odeur sucrée (limite olfactive : 5,4 ppm). Cette cétone est largement utilisée comme solvant entrant dans la composition des peintures, de vernis, d'encres, de colles et d'adhésifs (particulièrement les vinyliques, nitrocellulosiques ou acryliques). Elle sert également comme agent d'extraction de certaines huiles végétales et minérales [1]. La MEK peut être formée par décomposition anaérobique de la cellulose, de l'amidon, de l'hémicellulose et de la pectine par *Clostridium sp*. [2].

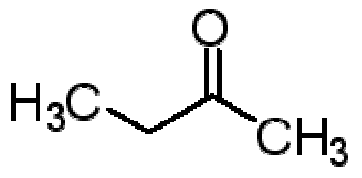

*Figure 1. Structure de la méthyléthylcétone (MEK).*

### 2. Principe de la photocatalyse hétérogène

La photocatalyse hétérogène est l'un des Procédés d'Oxydation Avancée (PAO) qui peut être réalisé à pressions et températures ambiantes sans ajout d'oxydants chimiques. La réaction photocatalytique est réalisée sur la surface d'un semi-conducteur selon les étapes suivantes : (1) production de paires d'électron ($e^-$)/lacune ($h^+$) par irradiation d'un semi-conducteur avec des photons ayant une énergie supérieure à sa bande interdite ; (2) séparation des électrons et des lacunes photogénérés dû au piégeage par les espèces adsorbées sur le semi-conducteur ; (3) réactions redox entre les électrons, les lacunes et les adsorbats ; et (4) désorption des produits de réaction et reconstruction de la surface du semi-conducteur. Généralement, les sous produits de la dégradation photocatalytique sont le dioxyde de carbone ($CO_2$) et l'eau ($H_2O$) [3]. Le dioxyde de titane ($TiO_2$) est le semi-conducteur le plus communément utilisé pour l'oxydation photocatalytique des COVs en phase gazeuse [4].

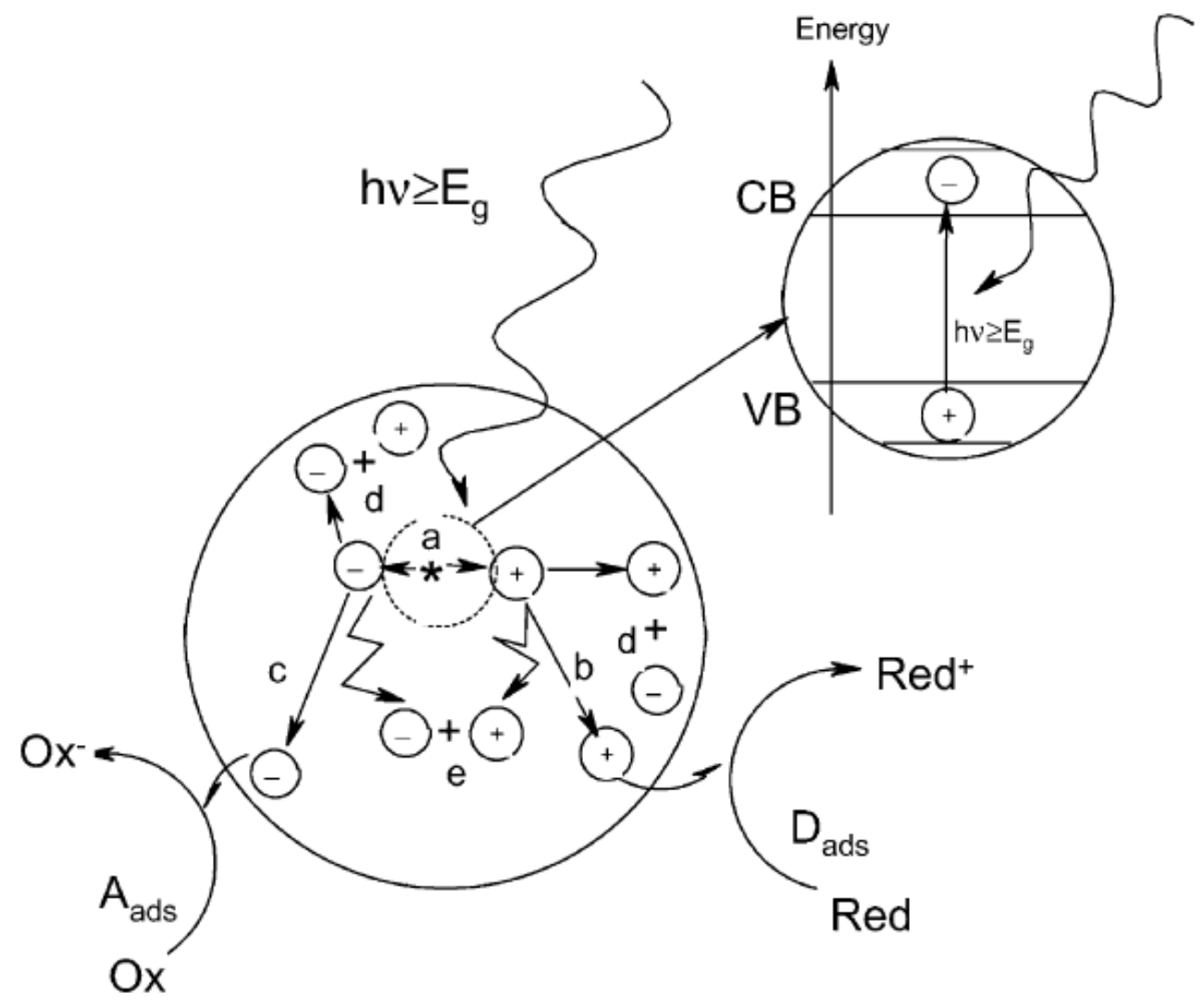

*Figure 2. Mécanisme de la photocatalyse hétérogène. (a) : génération des paires $e^-/h^+$ ; (b) : oxydation d'un donneur ($D_{ads}$) ; (c) : réduction d'un accepteur ($A_{ads}$) ; (d) et (e) : recombinaison des paires $e^-/h^+$.*

## DESCRIPTION DU DISPOSITIF EXPERIMENTAL

### 1. Montage expérimental

Le réacteur photocatalytique utilisé possède une géométrie annulaire et il est équipé de quatre entrées/quatre sorties pour assurer une meilleure distribution du flux gazeux (Figure 3).

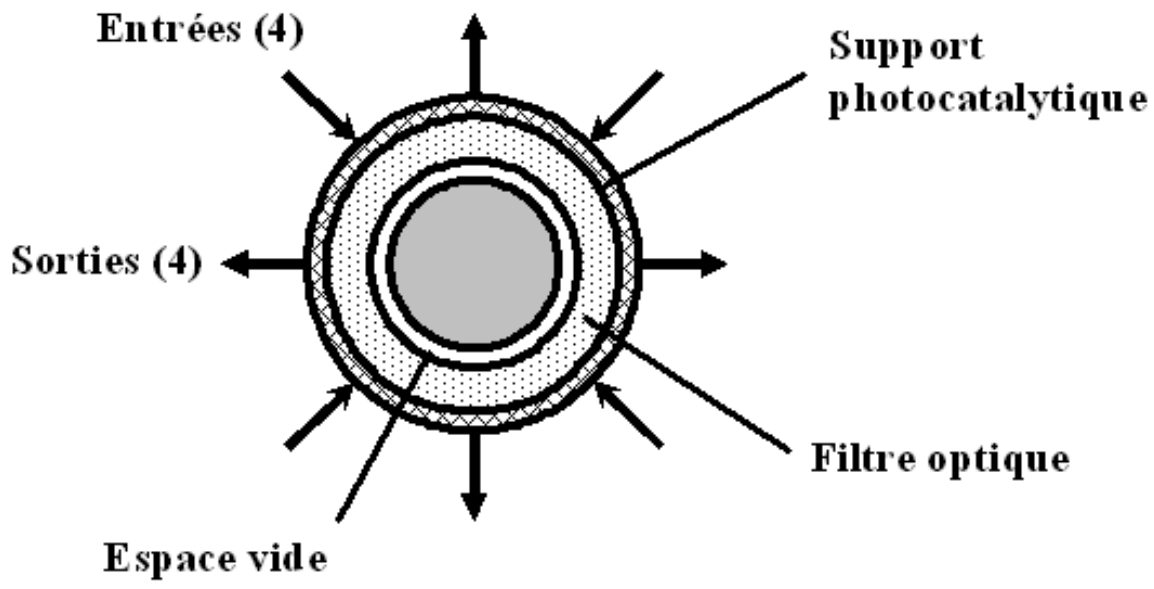

*Figure 3. Schéma du réacteur photocatalytique annulaire.*

Le catalyseur est inséré entre deux tubes en Pyrex afin d'obtenir le meilleur contact entre le catalyseur et l'air pollué. La position centrale du tube fluorescent offre les meilleures conditions d'éclairement. Le tube fluorescent et le catalyseur sont séparés par un filtre liquide pour contrôler en simultané la température et l'éclairement incident pendant l'oxydation photocatalytique. La transmission de la lumière est atténuée par l'utilisation de nigrosine en solution aqueuse dissoute directement dans le bain thermostaté (Figure 4). Le diamètre total du photoréacteur annulaire est de 52 mm et l'aire du support photocatalytique exposée aux UV est de 360 cm$^2$.

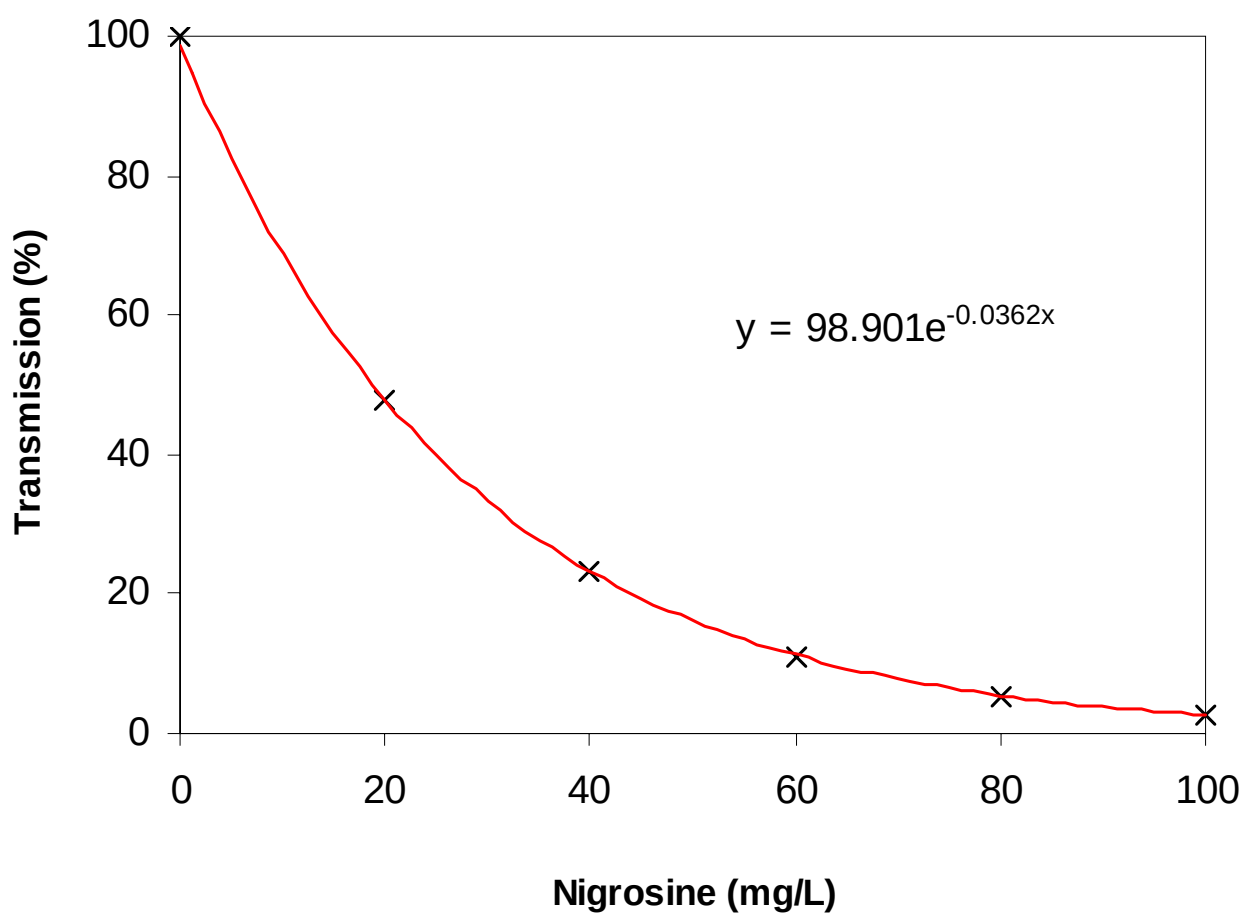

*Figure 4. Transmission lumineuse à travers différentes solutions de nigrosine à 365 nm.*

La source UV utilisé est un tube fluorescent (lumière noire) MAZDA 18 TWFN avec un pic spectral centré aux alentours de 365 nm. D'après la Figure 5, l'air est divisé en trois voies, chacune contrôlée par un débitmètre massique. Pour générer une atmosphère polluée, l'air barbote à travers deux saturateurs maintenus sous forte agitation dans un bain thermostaté, l'un contient le polluant et l'autre contient de l'eau. Le fonctionnement de ce dispositif expérimental a été largement détaillé dans une étude précédente [5].

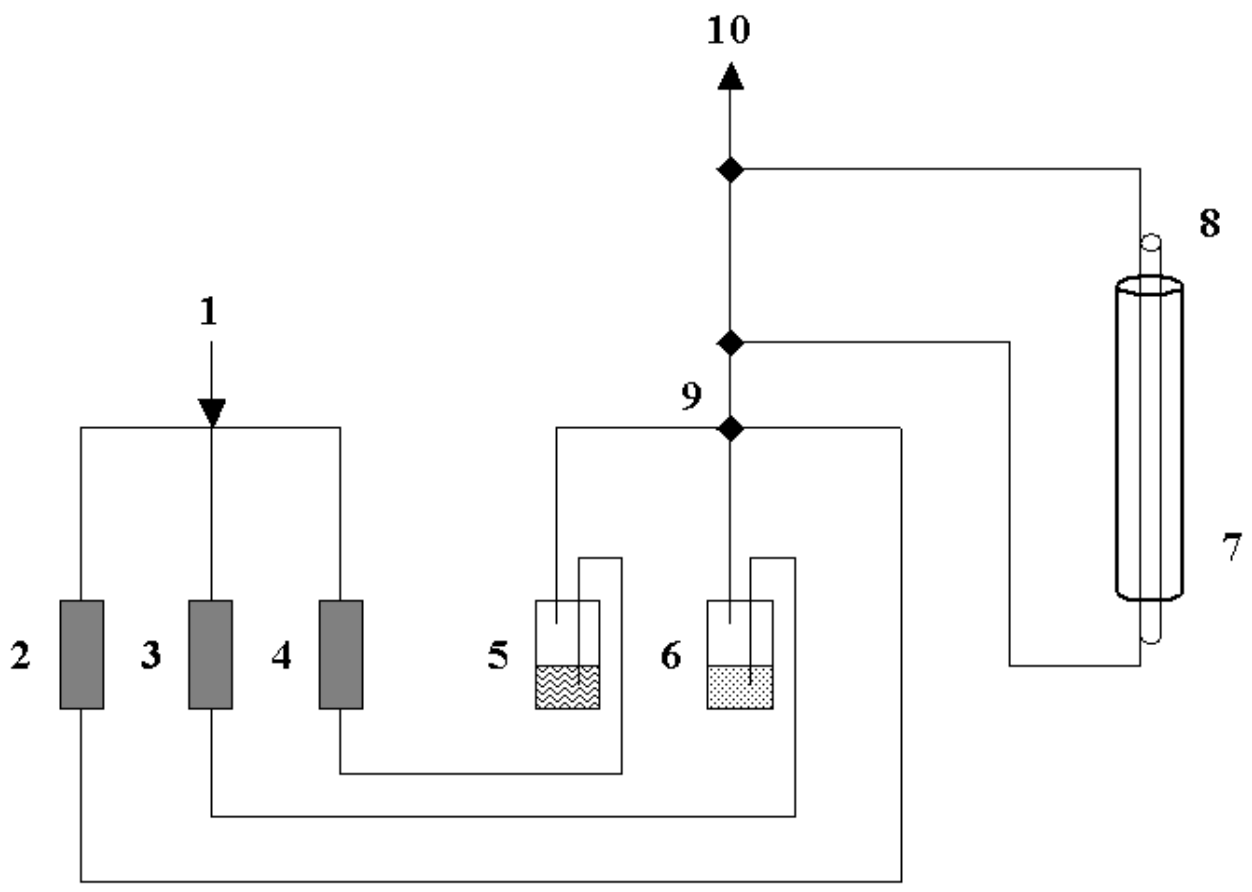

*Figure 5. Schéma du dispositif expérimental (1) arrivée d'air, (2) air sec, (3) air sec + $H_2O$, (4) air sec + COV, (5) saturateur COV, (6) saturateur $H_2O$, (7) réacteur photocatalytique, (8) tube fluorescent, (9) mélangeur statique, (10) chromatographe « gaz ».*

Un chromatographe « gaz » équipé d'un détecteur à ionisation de flamme (FID) a été utilisé pour suivre la concentration en MEK pendant la dégradation photocatalytique. La conversion $X$ de la MEK au sein du photoréacteur annulaire est donnée par la relation suivante :

$$X = 1 - \frac{A_{out}}{A_{in}}$$

Où $A_{out}$ (µV*s) représente l'aire du pic à la sortie du réacteur, $A_{in}$ (µV*s) l'aire du pic à l'entrée du réacteur.

## 2. Préparation du catalyseur

Lors de cette étude, nous avons utilisé du dioxyde de titane $TiO_2$ Degussa P25 disposé sur un support en fibres de verre (250 x 144 mm). Degussa P25 possède une surface spécifique d'environ 50 $m^2/g$ et une mixture de deux phases cristallines (70% anatase + 30% rutile). La déposition du catalyseur doit suivre un protocole strict détaillé dans une étude précédente [6]. Le dioxyde de titane Degussa P25 sous forme de poudre est dispersé dans une solution aqueuse en présence d'acide nitrique (pH = 3) qui prévient l'agrégation du dioxyde de titane pendant le mélange en solution. Ensuite, le support en fibres de verre est imprégné avec la suspension de dioxyde de titane. Après évaporation totale de l'eau, le support est séché à 100°C pendant 1h et calciné à 475°C pendant 4h afin d'assurer une bonne adhésion entre le catalyseur et le support. Environ 38 mg de $TiO_2$ ont été déposés sur le support en fibres de verre.

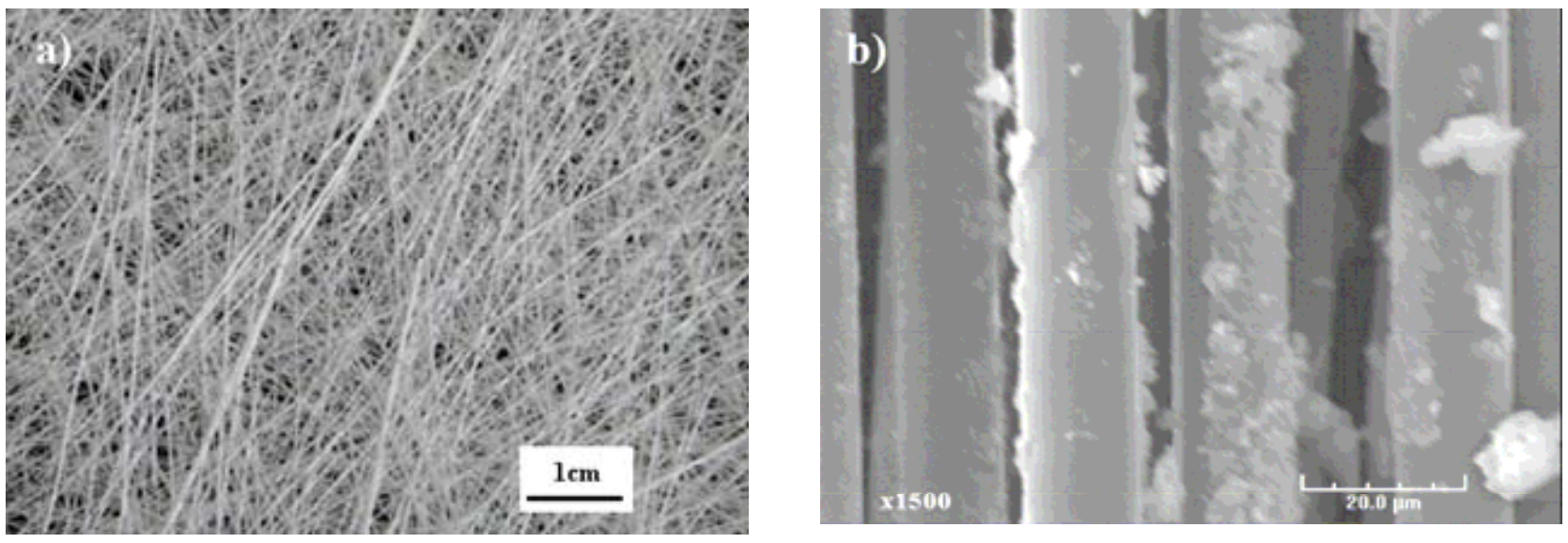

*Figure 6. a) support en fibres de verre, b) particules de $TiO_2$ imprégnées sur le support photocatalytique.*

# DISTRIBUTION DU TEMPS DE SEJOUR (DTS) AU SEIN DU PHOTOREACTEUR

## 1. Définition de la DTS

Toutes les molécules arrivant dans un réacteur ne transitent pas forcément de la même façon et elles peuvent rester plus ou moins longtemps avant d'en sortir. Le temps de séjour $t_s$ d'une molécule est défini comme le temps qu'elle passe au sein du réacteur. La DTS permet de caractériser le type de réacteur en présence : si les molécules possèdent toutes le même temps de séjour, le réacteur est piston et le signal $E(t_s)$ mesuré est un pic très étroit. Si les molécules ont des temps de séjour variables, on obtient un pic plus ou moins étalé et symétrique qui traduit l'écartement par rapport au cas idéal du piston. La DTS est donnée par l'équation suivante [7] :

$$E(t_s) = \frac{C(t_s)}{\int_0^{\infty} C(t_s)dt_s} = \frac{y(t_s)}{\sum_0^n y(t_s)\Delta t_s}$$

Où $E(t_s)dt_s$ représente la fraction de fluide ayant un temps de séjour compris entre $t_s$ et $t_s + dt_s$, $C(t_s)$ la concentration des molécules sortant du réacteur au temps $t_s$, $y(t_s)$ une grandeur mesurable, proportionnelle à $C(t_s)$, $\Delta t_s$ l'intervalle de temps entre chaque mesure, $n$ le nombre de mesures effectuées.

La détermination expérimentale de la DTS consiste à injecter une petite quantité d'un gaz traceur à l'entré du réacteur et de mesurer sa concentration dans le courant de sortie comme une fonction du temps.

## 2. Modèle des mélangeurs en cascade

On peut tenter de représenter l'écoulement du fluide dans un réacteur réel en assimilant celui-ci à une cascade de *J* réacteurs agités en série de même volume total (Figure 7).

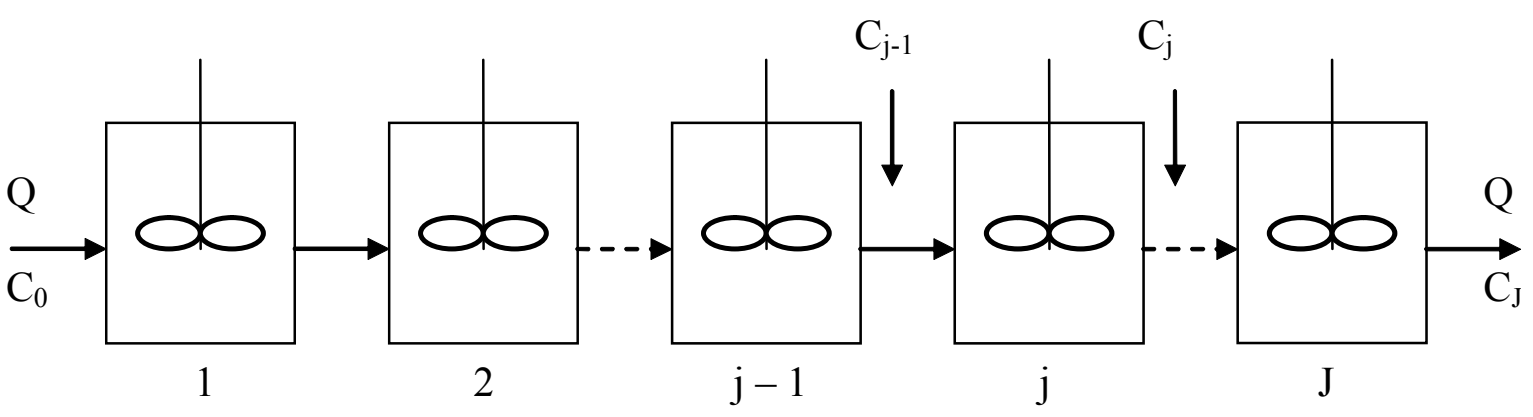

*Figure 7. Modèle des mélangeurs en cascade.*

Dans ce cas, la DTS est une distribution Gamma (ou de Poisson) :

$$E(t_s) = \left(\frac{J}{\overline{t_s}}\right)^J \frac{t_s^{\,J-1} \exp\left(-Jt_s / \overline{t_s}\right)}{\left(J-1\right)!}$$

Où *J* représente le nombre de réacteurs en cascade, $\overline{t_s}$ le temps de séjour moyen.

$J = 1$ correspond bien entendu au réacteur agité continu unique. Lorsque $J \rightarrow \infty$, on se rapproche de l'écoulement piston et les courbes de DTS sont voisines de gaussiennes symétriques. Un logiciel d'optimisation (Easyplot) permet de générer des valeurs de $J$ et $\overline{t_s}$ qui satisfont au mieux les points expérimentaux au sens des moindres carrés. Pour le photoréacteur annulaire, le logiciel d'optimisation a généré les valeurs suivantes : $J = 18$, $\overline{t_s} =$ 54,18 s. Le photoréacteur annulaire se rapproche du réacteur idéal piston car en pratique, on considère que le réacteur est piston lorsque $J > 20$.

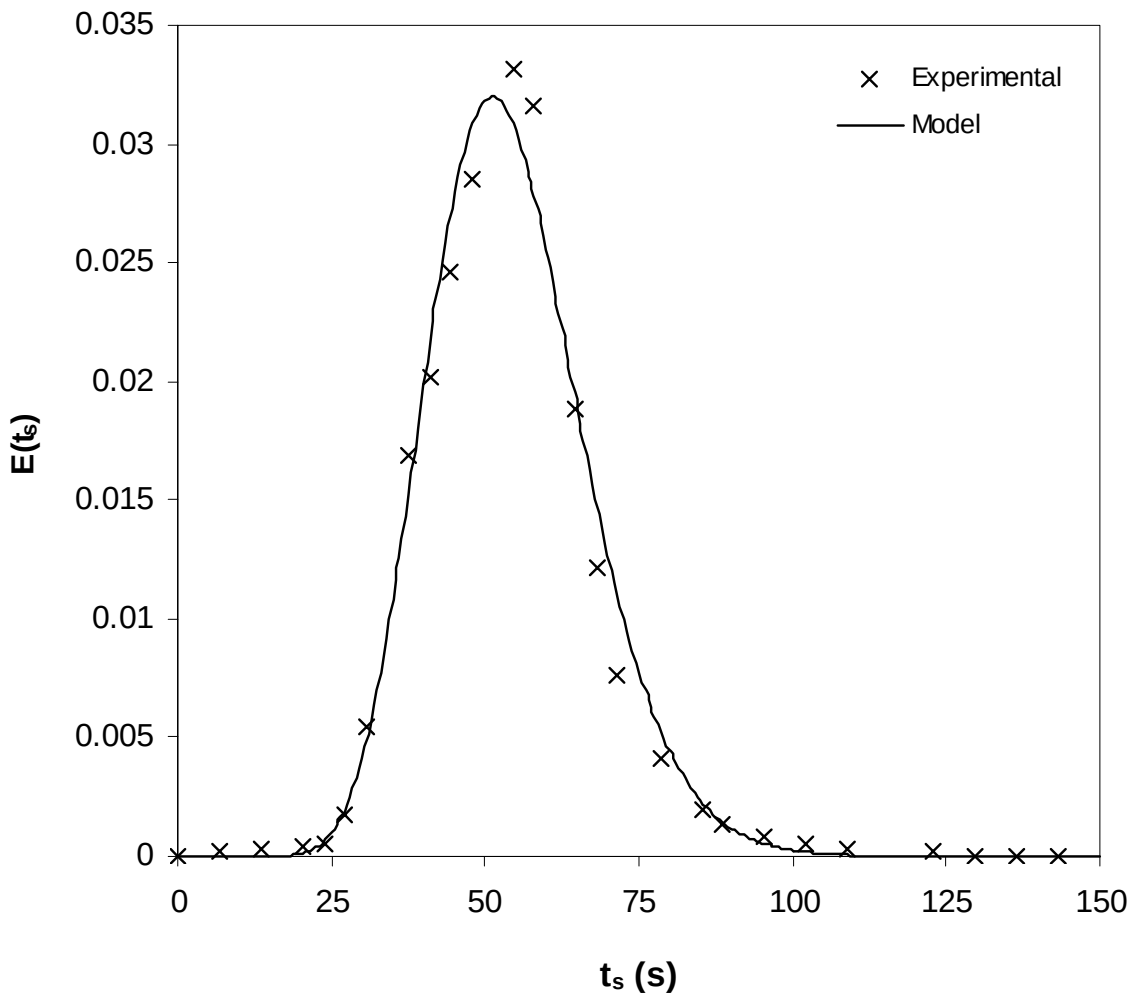

*Figure 8. Courbe de DTS obtenue pour le photoréacteur annulaire.*

# ETUDE CINETIQUE RELATIVE A LA METHYLETHYLCETONE (MEK)

## 1. Effet du transfert de masse

Dans les systèmes gaz/solide, un transfert de masse peut apparaître entre la phase gazeuse et la phase solide. Si l'influence du transfert de masse est significative, la vitesse de dégradation augmente avec le débit volumique. L'effet du transfert de masse a été étudié en utilisant différents débits volumiques allant de 100 à 340 mL/min en maintenant la concentration en MEK constante. Dans notre étude, la vitesse de dégradation est exprimée par unité de surface du support photocatalytique. Par conséquent, la vitesse de photodégradation de la MEK ($r_{MEK}$) est définie par l'expression suivante :

$$r_{MEK} = -\frac{d[MEK]_s}{d\left(\frac{S}{Q_v}\right)}$$

Où $[MEK]_s$ représente la concentration en MEK à la sortie du réacteur, $S$ la surface apparente du support photocatalytique, $Q_v$ le débit volumique.

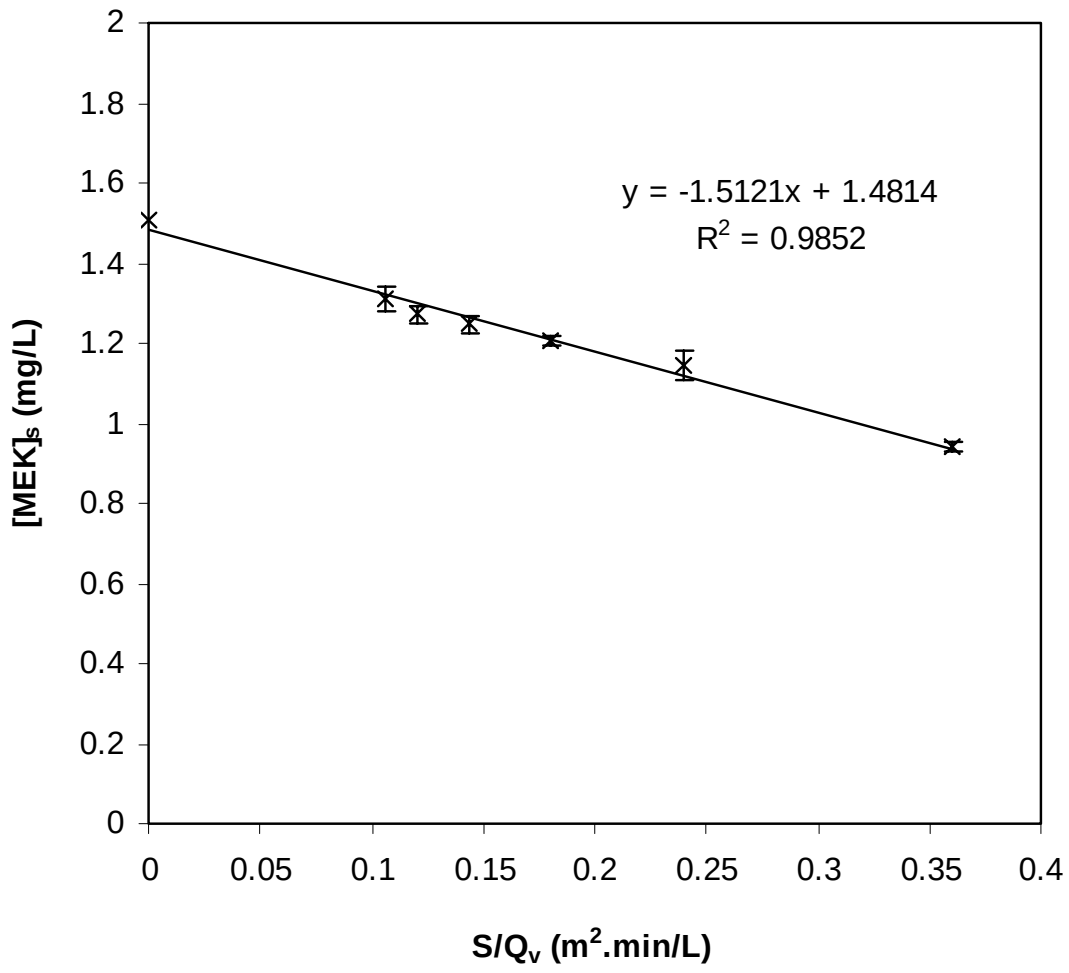

*Figure 9. Effet du débit volumique sur la vitesse de photodégradation.*

D'après la Figure 9, la vitesse apparente de disparition de la MEK est directement obtenue par la pente de la droite. On peut alors facilement voir que la vitesse de disparition de la MEK est constante quelque soit le débit volumique. Par conséquent, il semblerait que la vitesse de dégradation ne soit pas contrôlée par le transfert de matière.

## 2. Effet de la concentration initiale

Les mesures de DTS réalisées précédemment ont permis de déterminer le nombre J de réacteurs en cascade. Par conséquent, le bilan de matière correspondant à un réacteur de la cascade est exprimé par la relation suivante :

$$C_j = C_{j-1} - \varepsilon \frac{S}{JQ_v} \cdot r_j$$

Où $S$ représente la surface apparente du support photocatalytique, $\varepsilon$ le taux de vide, $J$ le nombre de réacteurs en cascade, $Q_v$ le débit volumique, $r_j$ la vitesse de réaction dans le réacteur de rang $j$.

La vitesse de dégradation de la méthyléthylcétone peut être représentée par le modèle de Langmuir-Hinshelwood (LH) [8]. La vitesse de réaction dans le réacteur de rang $j$ ($r_j$) est alors proportionnelle au taux de recouvrement moyen $\theta_j$ de la surface du catalyseur :

$$r_j = k\theta_j \quad \text{avec} \quad \theta_j = \frac{KC_j}{1+KC_j}$$

Où $k$ représente la constante cinétique de dégradation (mg/min/m$^2$), $K$ la constante d'équilibre d'adsorption (L/mg).

Ces constantes $k$ et $K$ peuvent être déterminées à partir du taux de conversion expérimental $X$ pour un débit donné et pour différentes valeurs de la concentration initiale $C_0$ à l'entrée de la cascade, en minimisant les différences quadratiques $\chi^2$ entres les valeurs expérimentales et théoriques des taux de conversion :

$$\chi^2 = \sum \left( X - X_J \right)^2 \quad \text{avec} \quad X_J = 1 - \frac{C_J}{C_0}$$

Où $X_J$ représente la conversion optimisée à partir des paramètres du modèle de LH, $X$ la conversion expérimental, $C_J$ la concentration optimisée en sortie de la cascade de réacteurs, $C_0$ la concentration initiale.

L'effet de la concentration initiale $[MEK]_0$ sur la vitesse de photodégradation a été étudié pour des concentrations allant de 0,094 à 1,503 mg/L avec un débit volumique constant de 300 mL/min. Après optimisation, les valeurs des constantes du modèle de LH sont les suivantes :
$k$ = 1,23 mg/min/m$^2$ et $K$ = 24,81 L/mg.

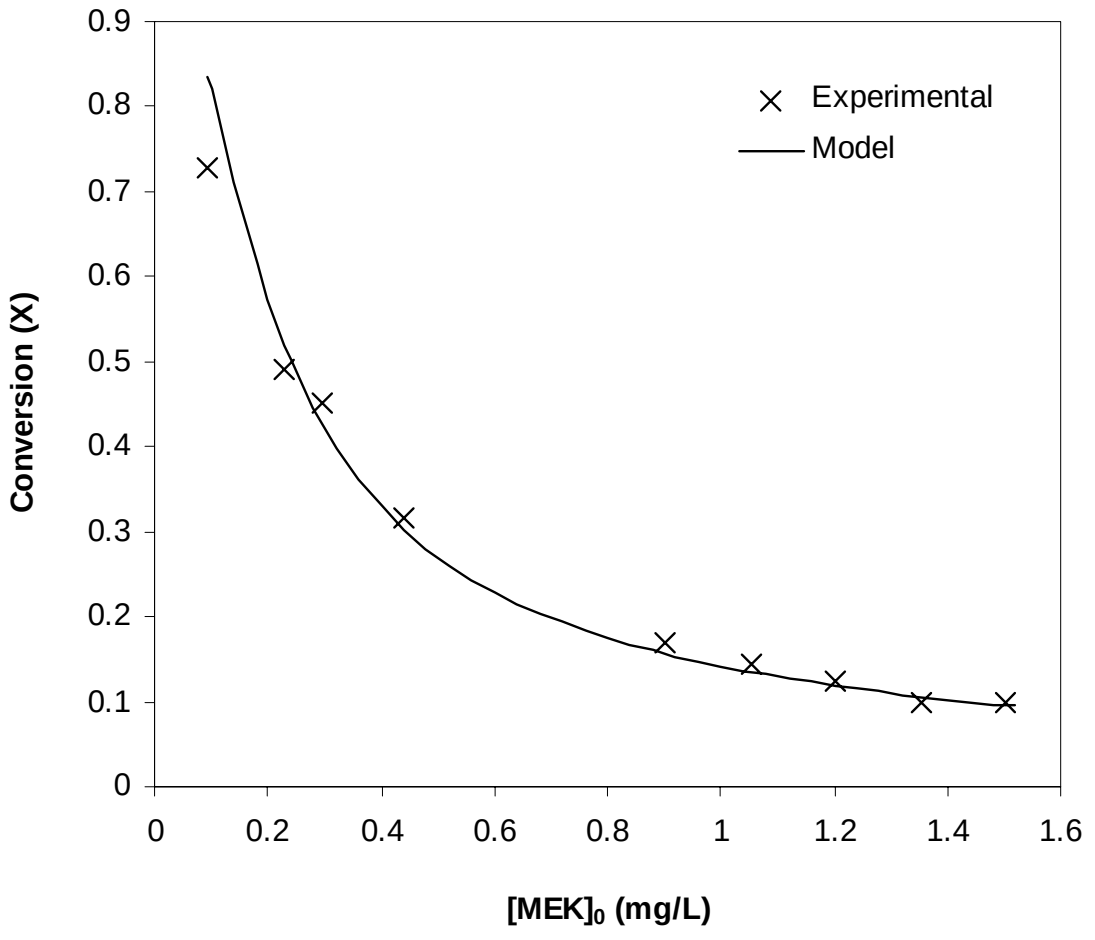

*Figure 10. Evolution de la conversion de la méthyléthylcétone en fonction de la concentration initiale.*

### 3. Effet de l'éclairement incident

L'effet de l'éclairement incident ($I_0$) sur la vitesse de photodegradation a été étudié pour des valeurs allant de 0,11 à 3,94 mW/cm$^2$. D'après Wang et al. [9], la constante cinétique de dégradation est fonction de l'éclairement incident :

$$k = k''\cdot I_0^n$$

Où *k''* est une constante indépendante de l'éclairement incident, *n* l'ordre cinétique compris entre $0 < n < 1$.

La vitesse de réaction dans le réacteur de rang *j* ($r_j$) est alors proportionnelle à l'éclairement incident atteignant la surface du catalyseur :

$$C_j = C_{j-1} - \varepsilon \frac{S}{JQ_v} \cdot k'' I_0^n \cdot \theta_{MEK}$$

Où $\theta_{MEK}$ représente le taux de recouvrement par la méthyléthylcétone à la surface du catalyseur.

Les valeurs des constantes *k''* et *n* ont été obtenues de la même manière que précédemment en minimisant les différences quadratiques $\chi^2$ entres les valeurs expérimentales et théoriques des taux de conversion. Après optimisation, nous pouvons conclure que la vitesse de dégradation suit une dépendance linéaire avec $I_0^{0,34}$.

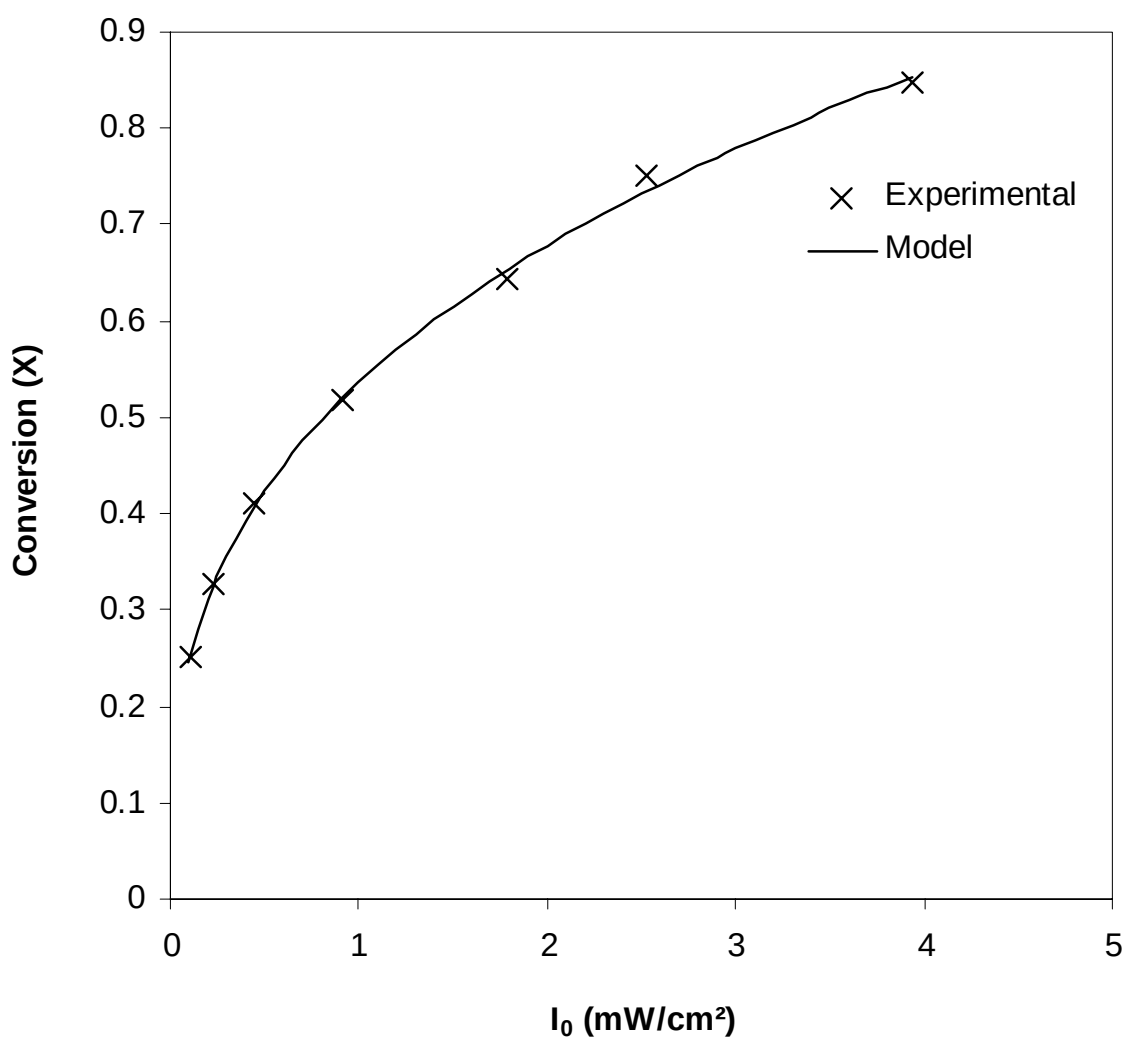

*Figure 11. Evolution de la conversion de la méthyléthylcétone en fonction de l'éclairement incident.*

D'après Ching et al. [10], trois cas peuvent se présenter :

- A faible éclairement, $n = 1$ : il existe une relation linéaire entre la vitesse de dégradation r et l'éclairement ($r$ α $I_0$). Cela signifie que les paires $e^-/h^+$ sont plus rapidement consommées par les réactions chimiques (réduction, oxydation) que par les phénomènes de recombinaison ;

- A éclairement intermédiaire, $n$ = 0,5 : le taux de formation des paires $e^-/h^+$ excède le taux d'oxydation catalytique entraînant une recombinaison des paires $e^-/h^+$ ($r$ α $I_0^{0,5}$) ;

- A fort éclairement, $n$ = 0 : l'éclairement n'a plus aucun effet sur la vitesse de dégradation. Dans ce cas, le photocatalyseur a atteint sa limite d'absorption de photons ou la vitesse de dégradation est limitée par le transfert de masse ($r$ α $I_0^0$).

Dans notre cas, la relation obtenue entre la vitesse de dégradation et l'éclairement suggère que la formation des paires $e^-/h^+$ excèdent la vitesse de dégradation entraînant une recombinaison des paires $e^-/h^+$ ($n$ = 0,34).

## 4. Identification des intermédiaires gazeux

D'après Raillard et al. [8], la dégradation photocatalytique du MEK entraîne la formation d'un ou plusieurs intermédiaires gazeux comme le méthyle formate et principalement l'acétaldéhyde. Nous avons tenté d'identifier les sous-produits de la photodégradation par Chromatographie en phase Gazeuse (Agilent® 6850 series GC system) couplée à un Spectromètre de Masse (Agilent® 5973 Network Mass Selective Detector).

D'après le chromatogramme, ci-dessous, nous avons pu identifier l'acétaldéhyde ($CH_3CHO$) comme intermédiaire gazeux majoritaire. L'acétaldéhyde et la 2-butanone (MEK) ont respectivement des temps de rétention de 15,22 et 20,11 min.

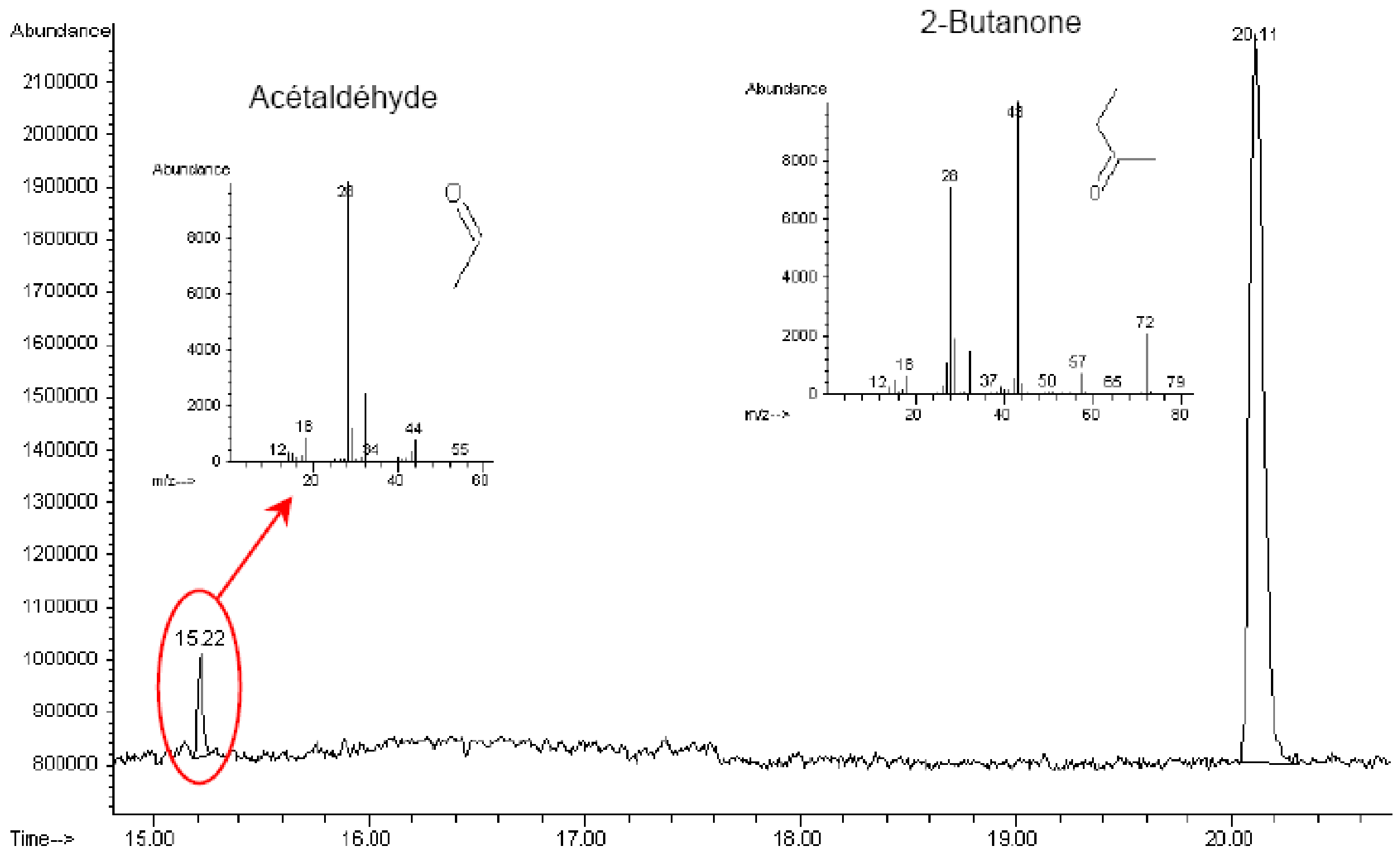

*Figure 12. Chromatogramme obtenu par GC/MS des intermédiaires gazeux de la dégradation photocatalytique de la MEK.*

Il est très important d'étudier l'influence de l'acétaldéhyde sur la vitesse de dégradation de la MEK. En effet, les intermédiaires générés durant le processus photocatalytique peuvent affecter la vitesse de dégradation en rentrant en compétition d'adsorption avec le polluant initial. Si nous considérons l'acétaldéhyde comme le seul intermédiaire majoritaire pendant le processus de photodégradation, le modèle cinétique de LH devra être réécrit de la manière suivante :

$$r = \frac{kKC}{1 + KC + K'C'}$$

Où *K'* représente la constante d'équilibre d'adsorption de l'acétaldéhyde, *C'* la concentration d'acétaldéhyde.

## CONCLUSIONS – PERSPECTIVES

Une dégradation efficace de la méthyléthylcétone a été observée sur le support en fibres de verre imprégné de $TiO_2$ Degussa P25. Le transfert de masse est négligeable sous les conditions expérimentales testées. En première approximation, la dégradation photocatalytique semble suivre comme la plupart des COVs le modèle cinétique de Langmuir-Hinshelwood. Cependant, l'acétaldéhyde considéré comme le principal intermédiaire gazeux devra être incorporé au modèle de LH pour montrer une éventuelle compétition d'adsorption entre la méthyléthylcétone et l'acétaldéhyde. Par la suite, un mécanisme de photodégradation de la MEK pourra être mis en évidence afin d'anticiper tous les intermédiaires gazeux susceptibles d'être produits pendant le processus d'oxydation photocatalytique de ce polluant.